\documentclass[11pt,a4paper]{article}
\usepackage{jheppub}
\usepackage{amsmath}
\usepackage[most]{tcolorbox}
\usepackage{dsfont}
\usepackage{ulem}
\usepackage{natbib}
\usepackage{xcolor}
\usepackage[hang,flushmargin]{footmisc}
\usepackage{tikz-cd}
\usepackage{enumitem}
\usepackage{amsfonts,amssymb, amscd,amsmath,latexsym,amsbsy,bm}
\usepackage{stmaryrd}
\usepackage{todonotes}
\usepackage{float}

\usepackage{romannum}

\makeatletter\renewcommand{\@biblabel}[1]{#1.}\makeatother

\newtcolorbox{empheqboxed}{colback=gray!20, 
 colframe=white,
 width=\textwidth,
 sharpish corners,
 top=0mm, 
 bottom=0pt
}

\title{The star-square relation and the generalized star-triangle relation from 3d supersymmetric dualities  \Romannum{1}}
\author{Mustafa Mullahasanoglu}

\affiliation{
	
{Department of Physics, Bogazici University, 34342 Bebek, Istanbul, Turkey}\\[-0.5cm]

}

\emailAdd{mustafa.mullahasanoglu@std.bogazici.edu.tr}

\abstract{
We study duality transformations of the star-square relation and the generalized star-triangle relation for Ising-like lattice spin models. The lattice spin models are obtained via gauge/YBE correspondence which connects the supersymmetric gauge theories and lattice spin models of statistical mechanics. By the use of integral identities coming from the duality of three-dimensional supersymmetric gauge theories, we construct hyperbolic, lens hyperbolic, trigonometric, and rational solutions to the duality transformations. These duality transformations allow us to construct spin lattice models with four-spin (the star-square relation) and three-spin (the generalized star-triangle relation) interactions.
}


\begin{document}

\maketitle

\section{Introduction}

The gauge/YBE correspondence \cite{Spiridonov:2010em, Yamazaki:2012cp} is one of the studies focusing on the links between supersymmetric gauge theories and integrable lattice spin models of statistical mechanics. The correspondence, see exhaustive reviews e.g \cite{Gahramanov:2017ysd, Yamazaki:2018xbx}, provides solutions to integrability conditions of lattice spin models the star-triangle relation \cite{Baxter:1982zz} and the star-star relation \cite{Baxter:1997tn} via the integral identities coming from the equality of partition functions of supersymmetric gauge theories. 

Along with various solutions to the integrability equations, the most general solution is obtained in terms of the lens elliptic gamma function \cite{Yamazaki:2013nra, Kels:2015bda} by using the duality of four-dimensional $\mathcal{N}=1$ supersymmetric gauge theories on $S_b^3/\mathbb{Z}_r \times S^1$. This lens elliptic model can be reduced to the lens hyperbolic model by the dimensional reduction and the star-triangle relation of this integrable model corresponds to the duality of three-dimensional $\mathcal{N}=2$ supersymmetric gauge theories on $S_b^3/\mathbb{Z}_r$ \cite{Gahramanov:2016ilb}. Besides the dimensional reduction, the gauge symmetry-breaking method from the gauge theory perspective also provides different integrable models. As an example, the generalized Faddeev-Volkov model \cite{Bozkurt:2020gyy, Mullahasanoglu:2021xyf} is acquired by breaking gauge symmetry $SU(2)$ to the $U(1)$ gauge group for the dualities of three-dimensional $\mathcal{N}=2$ supersymmetric gauge theories on $S_b^3/\mathbb{Z}_r$. The other known integrable models in this context are also obtained from the lens elliptic model \cite{Bazhanov:2010kz, Kels:2017vbc}.

The Ising-like models have spins sitting on sites and nearest-neighbor spins interact with each other through edges. However, one can also acquire the IRF-type (interaction round a face) models \cite{Yamazaki:2013nra, Yamazaki:2015voa, Gahramanov:2017idz, Mullahasanoglu:2021xyf}, and vertex-type (spins interacting at a vertex) \cite{Gahramanov:2015cva, Gahramanov:2022jxz, Catak:2022glx} integrable lattice spin models by using integrability conditions of edge interacting models and constructing Bailey pairs of them, respectively.

In this study, we investigate the non-planar lattice spin models consisting of higher-spin interactions by using the duality of the three-dimensional supersymmetric gauge theories. To achieve this investigation, we obtain solutions to the star-square relation \cite{Pais34, wegnerduality} and the generalized star-triangle relations \cite{Fisher1959, Stre_ka_2010} with the help of the gauge/YBE correspondence. 

This is the first time that higher spin interactions for Ising-like models are presented in the aspect of the gauge/YBE correspondence and Boltzmann weights of the dual higher spin interacting models are studied in terms of hyperbolic, lens hyperbolic, trigonometric, and rational functions.

The duality transformations -the star-square relation and the generalized star-triangle relation- equate partition functions of two different spin models up to some coefficient. In this duality of the lattice spin models, one model possesses only nearest-neighbor interactions and the dual model has various kinds of spin interactions such as higher spin\footnote{The term 'higher spin' is employed to keep the consistency with terminology in statistical mechanics and it should not be mistaken for its modern usage.} (triple or quadruple) interactions. From the supersymmetric gauge theory perspective, the corresponding solutions to the duality transformations are the integral identities resulting from the equality of the partition functions of the dual three-dimensional $\mathcal{N}=2$ supersymmetric gauge theories on $S_b^3$, $S_b^3/\mathbb{Z}_r$ and $S^2\times S^1$.
Therefore, Boltzmann weights of the models are written in terms of hyperbolic gamma function \cite{Faddeev:1994fw, Volkov:1992uv}, lens hyperbolic gamma function \cite{Gahramanov:2016ilb, Bozkurt:2020gyy, Mullahasanoglu:2021xyf}, trigonometric hypergeometric function \cite{Gahramanov:2016wxi, Gahramanov:2013rda, Gahramanov:2014ona}, Euler's gamma function \cite{Kels:2013ola, Sarkissian:2020ipg}.

In the future work \cite{Mullahasanoglu:202X}, we will apply the duality transformations for the integrable lattice spin models studied by the gauge/YBE correspondence, that is, the star-square relation and the generalized star-triangle relation will be solved in terms of Boltzmann weights belonging to the integrable models.

The paper is organized as follows. In Section 2, we briefly introduce all special functions used throughout the work. In the remaining two sections, we study the solutions to the star-square relation and the generalized star-triangle relation, respectively.

\section{Notations}


Let us start with introducing the $q$-Pochhammer symbol
\begin{align}
    (z;q)_{\infty}=\prod_{i=0}^{\infty}(1-zq^i)\:,
\end{align}
and the shorthand notation for the product of two $q$-Pochhammer symbols
\begin{align}
    (z,x;q)_\infty=(z;q)_\infty(x;q)_\infty\;.
\end{align}

The hyperbolic gamma function which is a variant of Faddeev's non-compact quantum dilogarithm \cite{van2007hyperbolic, Andersen:2014aoa} is another key special function that we utilize 
\begin{align}
	\gamma^{(2)}(z;\omega_{1},\omega_{2})=e^{\frac{\pi i}{2}B_{2,2}(z;\omega_{1},\omega_{2})}\frac{(e^{-2\pi i\frac{z}{\omega_{2}}}\tilde{q};\tilde{q})_\infty}{(e^{-2\pi i\frac{z}{\omega_{1}}};q)_\infty} \; ,
\end{align}
where  $\tilde{q}=e^{2\pi i \omega_{1}/\omega_{2}}$ and $q=e^{-2\pi i \omega_{2}/\omega_{1}}$ 
with the complex variables $\omega_{1}$, $\omega_{2}$  and with the second Bernoulli polynomial
\begin{equation}
 B_{2,2}(z;\omega_1,\omega_2)=\frac{z^2}{\omega_1\omega_2}-\frac{z}{\omega_1}-\frac{z}{\omega_2}+\frac{\omega_1}{6\omega_2}+\frac{\omega_2}{6\omega_1}+\frac{1}{2}\:.
\end{equation}

There are several integral representations for the hyperbolic gamma function, see, e.g. \cite{Faddeev:1995nb,woronowicz2000quantum}. We  give here one of them
\begin{align}
	\gamma^{(2)}(z;\omega_{1},\omega_{2})=\exp{\left(-\int_{0}^{\infty}\frac{dx}{x}\left[\frac{\sinh{x(2z-\omega_{1}-\omega_{2})}}{2\sinh{(x\omega_{1})}\sinh{(x\omega_{2})}}-\frac{2z-\omega_{1}-\omega_{2}}{2x\omega_{1}\omega_{2}}\right]\right)} \; ,
\end{align}
where $Re(\omega_{1}),Re(\omega_{2})>0$ and $Re(\omega_{1}+\omega_{2})>Re(z)>0$. 

The hyperbolic gamma function reduces to the Euler gamma function by the asymptotic property 
\begin{equation}
	\lim_{\omega_2\to\infty} \Big(\frac{\omega_2}{2\pi\omega_1}\Big)^{\frac{z}{\omega_2}-\frac{1}{2}}\gamma^{(2)}(z;\omega_1,\omega_2)=\frac{\Gamma(z/\omega_1)}{\sqrt{2\pi}}\:.
	\label{gamma_limit}
	\end{equation}
 
The hyperbolic gamma function has the reflection property 
\begin{align}
  \gamma^{(2)}(\omega_1+\omega_2- z ;\omega_1,\omega_2)\gamma^{(2)}( z ;\omega_1,\omega_2)=1 \:,\label{hypref}
\end{align}
and the shorthand notation is
\begin{align}
  \gamma^{(2)}(\pm z ;\omega_1,\omega_2)=\gamma^{(2)}(+ z ;\omega_1,\omega_2)\gamma^{(2)}(- z ;\omega_1,\omega_2) \:.
\end{align}
The reflection property is helpful while reducing the number of flavors in the supersymmetric gauge theories and the integral identities studied as the star-square relation will be reduced to obtain the generalized star-triangle relation. 

Lens hyperbolic gamma function \cite{Gahramanov:2016ilb} which is also related to the improved double sine function \cite{Nieri:2015yia} is defined with $\text{Im}(\omega_1/\omega_2)>0$ 
\begin{align}
\gamma_h(z,y;\omega_1,\omega_2) 
=    \gamma^{(2)}(-iz-i\omega_1y;-i\omega_1r,-i\omega) \times \gamma^{(2)}(-iz-i\omega_2(r-y);-i\omega_2r,-i\omega) \:,
\end{align}
where $r\in\{1,2,...\}$, $y\in \{0,1,...,r-1\}$, and  $\omega:=\omega_1+\omega_2$. The reflection property is 
\begin{align}
\gamma_h(\omega_1+\omega_2-z,r-y;\omega_1,\omega_2)\gamma_h(z,y;\omega_1,\omega_2) 
=    1\:.
\end{align}

As studied in \cite{Sarkissian:2020ipg}, we define the gamma function as a specific ratio of Euler's gamma functions

\begin{equation}
{\bf \Gamma}(x,n)
=\frac{\Gamma\big(\frac{n+{i}x}{2}\big)}{\Gamma\big(1+\frac{n-{ i}x}{2}\big)},
\label{Cgamma}\end{equation}
where $x\in {\mathbb C}$ and $n\in {\mathbb Z}$.
The reflection relation of the Euler gamma function $\Gamma(x)\Gamma(1-x)=\pi/\sin\pi x$ leads
\begin{equation}
{\bf \Gamma}(x,-n)=(-1)^n{\bf \Gamma}(x,n)\:,
\quad \&\quad
{\bf \Gamma}(x,n){\bf \Gamma}(-x-2{ i},n)=1\:.
\label{gamma}\end{equation}

\section{The star-square relation}

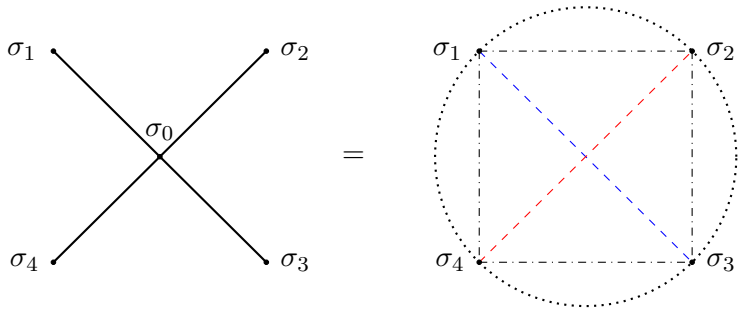
\begin{figure}[tbh]
\centering
\begin{tikzpicture}[scale=0.7]

\draw[-,thick] (-4,2)--(0,-2);
\draw[-,thick] (-4,-2)--(0,2);

\filldraw[fill=black,draw=black] (-2,0) circle (1.2pt)
node[above=2.5pt]{\color{black} $\sigma_0$};
\filldraw[fill=black,draw=black] (-4,2) circle (1.2pt)
node[left=2.5pt]{\color{black} $\sigma_1$};
\filldraw[fill=black,draw=black] (0,-2) circle (1.2pt)
node[right=1.5pt] {\color{black} $\sigma_3$};
\filldraw[fill=black,draw=black] (-4,-2) circle (1.2pt)
node[left=1.5pt] {\color{black} $\sigma_4$};
\filldraw[fill=black,draw=black] (0,2) circle (1.2pt)
node[right=1.5pt] {\color{black} $\sigma_2$};

\fill[white!] (2.05,0) circle (0.01pt)
node[left=0.05pt] {\color{black}$=$};

\draw[-,dashed, blue] (4,2)--(8,-2);
\draw[-,dashed, red] (4,-2)--(8,2);

\draw[-,dash dot] (4,2)--(4,-2);
\draw[-,dash dot] (4,2)--(8,2);
\draw[-,dash dot] (8,2)--(8,-2);
\draw[-,dash dot] (4,-2)--(8,-2);

\draw[dotted, thick] (6,0) circle (2.83cm);

\filldraw[fill=black,draw=black] (4,2) circle (1.2pt)
node[left=2.5pt]{\color{black} $\sigma_1$};
\filldraw[fill=black,draw=black] (8,-2) circle (1.2pt)
node[right=1.5pt] {\color{black} $\sigma_3$};
\filldraw[fill=black,draw=black] (4,-2) circle (1.2pt)
node[left=1.5pt] {\color{black} $\sigma_4$};
\filldraw[fill=black,draw=black] (8,2) circle (1.2pt)
node[right=1.5pt] {\color{black} $\sigma_2$};
\end{tikzpicture}
\caption{The star-square relation.}
\label{SSR}
\end{figure}

In this section, we focus on the star-square relation as shown in Fig.\ref{SSR} which is firstly discussed in  \cite{Pais34} and then studied from different perspectives\footnote{The star-square relation is used to work on some quantities such as universality and critical exponents \cite{Jungling_1974, Jungling_1975, Jungling_1976} in the context of the equivalence of Ising model to the 8-vertex model \cite{Wu1971, Baxter1971}, and it is also studied on the field of the differential renormalization group theory \cite{zewski:1981,vansaarlos}.} \cite{wegnerduality, Jungling_1975, vansaarlos}. 
The star-square relation represents the duality between two different kinds of lattice spin models in statistical mechanics and equates the partition functions of the models up to some coefficient. The star-square relation depicted in Fig.\ref{SSR}, consists of four nearest-neighbor interactions on the star side and seven interactions which are four nearest-neighbors (dashed and dotted lines), two diagonal-neighbors (dashed lines), and one quadruple (broken circle) interactions at the square side.

The star-square relation can be written as the following equation in which star side has only next-neighbor spin interactions $W_{\alpha,\beta}(\sigma_i,\sigma_j)$  and the other one has four-spin interactions $\widetilde{V}_{\sum_{i=1}^4\alpha_i,\sum_{i=1}^4\beta_i}(\sigma_1,\sigma_2,\sigma_3,\sigma_4)$ and diagonal-neighbor and nearest-neighbor two-spin interactions $V_{\alpha_i+\alpha_j,\beta_i+\beta_j}(\sigma_i,\sigma_j)$ which differ from previous edge interactions
\begin{align}
\sum_{v_0}  \int dx_0 \: S(\sigma_0)\:&\prod_{i=1}^4 \:W_{\alpha_i,\beta_i}(\sigma_i,\sigma_0)   
\nonumber \\
&  =
\widetilde{V}_{\sum_{i=1}^4\alpha_i,\sum_{i=1}^4\beta_i}(\sigma_1,\sigma_2,\sigma_3,\sigma_4)
   \prod_{1\leq i<j\leq 4} 
   V_{\alpha_i+\alpha_j,\beta_i+\beta_j}(\sigma_i,\sigma_j)     
  \:,\label{stsqr}
\end{align}
where $\sigma=(x,v)$'s are the spin variables representing both continuous $x$ and discrete $v$ spin variables and $\alpha$'s and $\beta$'s  are the spectral parameters. 

In this study, the Boltzmann weights represented as $W_{\alpha,\beta}(\sigma_i,\sigma_j)$ belong to planar models. However, the Boltzmann weights $V_{\alpha_i+\alpha_j,\beta_i+\beta_j}(\sigma_i,\sigma_j)$ of the dual non-planar lattice spin models are yet unknown whether they solve the integrability conditions without different types of Boltzmann weights. However, both Boltzmann weights asymmetrically solve the star-triangle relation seen for various cases in terms of hyperbolic gamma functions \cite{Kels:2018xge}. 

The star-square relation is used for the vertex formulation \cite{Sergeev:1995rt}
of the Bazhanov–Baxter model \cite{Bazhanov1992} to prove the tetrahedron equation \cite{Kashaev:1992tu}, see for review \cite{Stroganov1997}, and it is used as one of the main tools\footnote{Another tool is the inversion relation, see \cite{Gahramanov:2022qge, Spiridonov:2014cxa, Gahramanov:2023lwk, Bazhanov:2011mz, Bazhanov:2013bh} in the context of the gauge/YBE correspondence.} for the construction of three-dimensional integrable lattice spin models \cite{Mangazeev:1994, Boos:1995}, see also \cite{Hu1994, HU1995151}.



\subsection{Hyperbolic model}

Here we will discuss the integral identity  obtained in \cite{Kels:2018xge, Sarkissian:2020ipg} as a hyperbolic beta integral of Askey-Wilson type \cite{STOKMAN2005119, Ruijsenaars2003}
\begin{align}
    \int_{-i\infty}^{i\infty}&\frac{\prod_{j=1}^4\gamma^{(2)}(a_j\pm z;\omega_1,\omega_2)}{\gamma^{(2)}(\pm2z;\omega_1,\omega_2)}\frac{dz}{2i\sqrt{\omega_1\omega_2}}
    \nonumber\\
   & = \gamma^{(2)}(\omega_1+\omega_2-\sum_{i=1}^4a_i;\omega_1,\omega_2)
    \prod_{1\leq i<j\leq 4}\gamma^{(2)}(a_i+a_j;\omega_1,\omega_2) \:.
    \label{intssr}
\end{align}
Note that there is no balancing condition in (\ref{intssr}) and in the integral identities studied in the rest of the paper. Also, the mathematical structures of the integral identities, which we will not go into detail, can be found in \cite{BultThesis} and references therein.

From a statistical mechanics point of view, it first appeared as the solution to the star-triangle relation in \cite{Bazhanov:2015gra, Kels:2018xge}. We will show that this integral identity can be also written as the solution to the star-square relation.  

We would like to mention that one can obtain the integral identity (\ref{intssr}) by reducing the number of flavors from $N_f=6$ to $N_f=4$ in the equality of partition functions of a certain dual three-dimensional $\mathcal{N} = 2$ theories on the squashed three-sphere. Namely, one can use the duality between $SU(2)$ electric theory with $N_f=6$  flavors where chiral multiplets transform under the fundamental representation of the gauge group and the flavor group and the vector multiplet transforms as the adjoint representation of the gauge group, and the magnetic theory consisting of fifteen chiral multiples in the totally antisymmetric tensor representation of the flavor group, without gauge degrees of freedom. This duality is studied as the star-triangle relation in \cite{Spiridonov:2010em} and it is shown that the Faddeev-Volkov model can be obtained by breaking the gauge symmetry from $SU(2)$ to $U(1)$.

We will make the following change of the variables into (\ref{intssr})
\begin{align}
    \begin{array}{c}
 a_j=x_j-i\alpha_j\,,\quad 
  j=1,2,3,4\:,
\end{array}
\label{chngvs}
\end{align}
where $\alpha_i$ is introduced as a spectral parameter, $x_i$ is a continuous spin variable and $z$ will be denoted as central spin $x_0$ as in Fig.\ref{SSR} in the rest of the paper.

Then we solve the star-square relation  (\ref{stsqr}) by introducing the following Boltzmann weights
 \begin{align}
    \begin{aligned}
W_\alpha(x_i,x_j)=\gamma^{(2)}(-i\alpha +x_i\pm x_j ;\omega_1,\omega_2)\:,
\end{aligned}\label{integrableB1}
\end{align}
and Boltzmann weights of the dual non-planar lattice spin model
 \begin{align}
    \begin{aligned}
        V_{\alpha_i+\alpha_j}(x_i,x_j)=\gamma^{(2)}(-i(\alpha_i+\alpha_j) +x_i+ x_j ;\omega_1,\omega_2)\:,
        \\
        \widetilde{V}_{\sum_{i=1}^4\alpha_i}(x_1, x_2, x_3, x_4)=\gamma^{(2)}\left(\omega_1+\omega_2+i\sum_{i=1}^4\alpha_i\ -\sum_{i=1}^4x_i ;\omega_1,\omega_2\right) \:,
    \end{aligned}\label{dual1}
\end{align}
where $\omega_1, \omega_2$ are temperature-like parameters. The self-interaction term as an external field is
\begin{align}
   S(x_0)= \frac{1}{\gamma^{(2)}(\pm2x_0;\omega_1,\omega_2)}\:.\label{self1}
\end{align}

The solution with Boltzmann weights (\ref{integrableB1}) and (\ref{dual1}) reveals a duality (equality of partition functions up to some coefficient) between a planar lattice spin model
and a non-planar lattice spin model consisting of neighbor interactions, diagonal interactions, and four-spin interactions as presented in Fig.\ref{SSR}.

\subsection{(Lens) Hyperbolic model}

In this part, we study the equality of partition functions of three-dimensional $\mathcal N=2$ theories on the squashed lens space $S^3_b/\mathbb{Z}_r$. The equality of partition functions with $N_f=6$ flavors obtained via dimensional reduction \cite{Benini:2011nc, Yamazaki:2013fva, Eren:2019ibl} or localization techniques \cite{Imamura:2012rq, Imamura:2013qxa} is interpreted as star-triangle relation in \cite{Sarkissian:2018ppc, Gahramanov:2016ilb}. 

Here, we investigate the integral identity for the $N_f=4$ case which is obtained in \cite{Mullahasanoglu:2021xyf} by reduction of the number of flavors
\begin{align} \nonumber
    &\frac{1}{2r\sqrt{-\omega_1\omega_2}} \sum_{y=0}^{[ r/2 ]}\epsilon (y) \int _{-\infty}^{\infty} \frac{\prod_{i=1}^4\gamma_h(a_i\pm z,u_i\pm y;\omega_1,\omega_2)}{\gamma_h(\pm 2z,\pm 2y;\omega_1,\omega_2)}  dz
     \nonumber \\
     &=
          \gamma_h\left(\omega_1+\omega_2-\sum_{i=1}^4a_i,-\sum_{i=1}^4u_i;\omega_1,\omega_2\right)
          \prod_{1\leq i<j\leq 4}\gamma_h(a_i + a_j,u_i + u_j;\omega_1,\omega_2)  \; ,
          \label{intssr2}
\end{align}
where the function $\epsilon(y)$ is $\epsilon(0)=\epsilon(\lfloor\frac{r}{2}\rfloor)=1$, and for all other values of $y$, $\epsilon(y)=2$, the summation is over the holonomies $y=\frac{r}{2\pi} \int A_\mu d x^{\mu}$, where $A_\mu$ is the gauge field, and the integration is performed over a non-trivial cycle on $S_b^3/{\mathbb Z}_r$.

One can obtain the integral identity (\ref{intssr}) by fixing $r=1$ in (\ref{intssr2}) and so one can also study the integral identity (\ref{intssr2}) as a generalized version of the solution to the star-triangle relation studied in \cite{Kels:2018xge}. Also, the integral identity (\ref{intssr2}) is recently discussed as a solution to the decoration transformation \cite{Catak:2024ygo}. 

The change of variables can be done for both continuous variables $a_i$ and discrete variables $u_i$ with two types of spectral parameters \cite{Spiridonov:2019uuw, Gahramanov:2022jxz}
\begin{align}
    \begin{array}{c}
 a_j=x_j-i\alpha_j\,,\quad 
  u_j=v_j-i\beta_j\,,\quad j=1,2,3,4\:,
\end{array}
\label{chngvs2}
\end{align}
where $\beta_i$ and $v_i$ are discrete spectral parameters and discrete spin variables, respectively and we will denote $y$ as $v_0$.

We introduce the Boltzmann weights for the nearest-neighbor interactions of the planar lattice spin model 
\begin{align}
    \begin{aligned}
        W_{\alpha, \beta}(\sigma_i,\sigma_0)=\gamma_h(-i\alpha +x_i\pm x_0,-i\beta +v_i\pm v_0 ;\omega_1,\omega_2)\:,
        \label{integrableB2}
    \end{aligned}
\end{align}
and the Boltzmann weights of the dual model consisting of higher-spin and diagonal interactions to solve the star-square relation (\ref{stsqr}) 
\begin{align}
    \begin{aligned}
        V_{\alpha_i+\alpha_j, \beta_i+\beta_j}(\sigma_i,\sigma_j)=\gamma_h(-i(\alpha_i+\alpha_j) +x_i+ x_j,-i(\beta_i+\beta_j) +v_i+ v_j ;\omega_1,\omega_2)\:,
        \\
        \widetilde{V}_{\sum_{i=1}^4\alpha_i,\sum_{i=1}^4\beta_i}(\sigma_1,\sigma_2,\sigma_3,\sigma_4)=\gamma_h\left(\omega_1+\omega_2+\sum_{i=1}^4(i\alpha_i -x_i) ,\sum_{i=1}^4(i\beta_i-v_i) ;\omega_1,\omega_2\right) \:,\label{dual2}
    \end{aligned}
\end{align}
where we remember that $\sigma=(x,v)$ represents both continuous and discrete spin variables, and the self-interaction term is
\begin{align}
    S(\sigma_0)=\frac{1}{\gamma_h(\pm 2x_0,\pm 2v_0;\omega_1,\omega_2)}\:.\label{self2}
\end{align}

We obtain a dual model (\ref{dual2}) to the planar lattice spin model (\ref{integrableB2}) by using the star-square relation. The non-planar dual model has not only neighbor interactions but also diagonal interactions, and four-spin interactions.

\subsection{Trigonometric model}

From the gauge/YBE perspective, basic hypergeometric identities \cite{Gahramanov:2016wxi, Gahramanov:2013rda, Gahramanov:2014ona} are discussed as trigonometric solutions to the star-triangle relation and the star-star relation \cite{Gahramanov:2015cva, Catak:2021coz}. Similarly, a new solution to the star-square relation can be provided via the equality of partition functions on $S^2 \times S^1$ which represents the equivalence of the superconformal indices for dual theories \cite{Gahramanov:2013xsa, Gahramanov:2016wxi} in which the integral identity is

\begin{align} \nonumber
& \sum_{y\in\mathbb{Z}}\oint \frac{dz}{4 \pi i z}\frac{(1-q^{y} z^2)(1-q^{y} z^{-2})}{q^y     z^{4y} }
\prod_{j=1}^4 
\frac{(q^{1+\frac{u_j+y}{2} }/{a_jz},q^{1+\frac{u_j-y}{2} }{z}/{a_j};q)_\infty}
{(q^{\frac{u_j+y}{2} }a_jz,q^{\frac{u_j-y}{2} }{a_j}/{z};q)_\infty}
  \\ 
&\quad= \frac{( q^{\frac{\sum_{i=1}^4 u_i}{2}}a_1 a_2 a_3 a_4;q)_{\infty}}{(q^{1+ \frac{\sum_{i=1}^4 u_i}{2}}/a_1 a_2 a_3 a_4;q)_{\infty}} \prod_{1\leq j<k\leq 4}  \frac{(q^{1+\frac{u_j+u_k}{2}}/a_ja_k;q)_\infty}
{(q^{\frac{u_j+u_k}{2}}a_ja_k;q)_\infty}\:.
\label{s2s1int}
\end{align}

If one redefine fugacities as $a_j=\alpha_j^{-1}x_{j} $ and $u_j=v_j-\beta_j$, the Boltzmann weights for the lattice spin model are
\begin{align}
 \begin{aligned}
    W_{\alpha, \beta}(\sigma_i,\sigma_0)=
    \frac{(q^{1+(-\beta_i+v_i-v_0)/2}(\alpha_i^{-1}x_ix_0)^{-1},q^{1+(\beta_i+v_0-v_i)/2}(\alpha_i^{-1}x_ix_0^{-1})^{-1};q)_\infty}{(q^{(-\beta_i+v_i-v_0)/2}\alpha_i^{-1} x_ix_0,q^{(\beta_i+v_0-v_i)/2}\alpha^{-1}x_ix_0^{-1};q)_\infty}\;,
    \end{aligned}\label{integrableB3}
\end{align}
and the Boltzmann weights for the corresponding dual model take the following form
\begin{align}
 \begin{aligned}
 \widetilde{V}_{\sum_{i=1}^4\alpha_i,\sum_{i=1}^4\beta_i}(\sigma_1,\sigma_2,\sigma_3,\sigma_4)&=
        \frac{( q^{\frac{\sum_{i=1}^4(-\beta_i+v_i)}{2}}\prod_{i=1}^4\alpha_i^{-1}x_i;q)_{\infty}}{(q^{1+ \frac{\sum_{i=1}^4(-\beta_i+v_i)}{2}}\prod_{i=1}^4(\alpha_i^{-1} x_i)^{-1};q)_{\infty}} \:,
        \\
        V_{\alpha_i+\alpha_j, \beta_i+\beta_j}(\sigma_i,\sigma_j)&=\frac{(q^{1+\frac{\beta_i+\beta_j +v_i+ v_j}{2}}(\alpha_i^{-1}\alpha_j^{-1} x_i x_j)^{-1};q)_\infty}
{(q^{\frac{\beta_i+\beta_j +v_i+ v_j}{2}}\alpha_i^{-1}\alpha_j^{-1} x_i x_j;q)_\infty}
        \:.
    \end{aligned}\label{dual3}
\end{align}

The self-interaction term appears due to the external field on the integrable model is
\begin{align}
    S(\sigma_0)=\frac{(1-q^{v_0} x_0^2)(1-q^{v_0} x_0^{-2})}{q^{v_0}     x_0^{4v_0} }\:.\label{self3}
\end{align}

The planar lattice spin model with nearest neighbor interactions is dual to the non-planar lattice spin model via the star-square relation as depicted in Fig.\ref{SSR}. 

\subsection{Rational model}
The rational solution to the star-square relation can be obtained from the lens hyperbolic solution (\ref{intssr2}) by applying the limit $r\to \infty$ using the asymptotic property of the lens gamma function (\ref{gamma_limit}). 
The integral identity is also investigated in \cite{Sarkissian:2020ipg}
as a general complex analogue of the de Branges-Wilson integral \cite{andrews_askey_roy_1999} and see for particular cases \cite{Derkachov_2020,Neretin_2020}
\begin{align}\begin{aligned}
\sum_{y\in {\mathbb Z}}&\int_{-\infty}^{\infty}\big(z^2+y^2\big)\prod_{i=1}^4{\bf \Gamma}(a_i\pm z,u_i\pm y)\frac{dz}{ 8\pi}
\\
&={\bf \Gamma}\Big(-2i-\sum\limits_{i=1}^ 4 a_i,\sum\limits_{i=1}^ 4 u_i\Big)
\prod\limits_{1\leq i < j \leq 4}
{\bf \Gamma}(a_i+a_j,u_i+u_j)\:.
\label{hyper44}
\end{aligned}\end{align} 

We study the solution of the star-square relation to acquire a dual of the edge interaction model in terms of rational functions. We apply the change of variables (\ref{chngvs2}) and obtain the following Boltzmann weights of a lattice spin model
 \begin{align}
    \begin{aligned}
        W_{\alpha, \beta}(\sigma_i,\sigma_0)={\bf \Gamma}(-i\alpha +x_i\pm x_0,-i\beta +v_i\pm v_0 )
       \:, \label{integrableB4}
    \end{aligned}
\end{align}
and Boltzmann weights of dual lattice spin model
 \begin{align}
    \begin{aligned}
    \widetilde{V}_{\sum_{i=1}^4\alpha_i,\sum_{i=1}^4\beta_i}(\sigma_1,\sigma_2,\sigma_3,\sigma_4)&={\bf \Gamma}\left(-2i-\sum_{i=1}^4(-i\alpha_i +x_i) ,\sum_{i=1}^4(-i\beta_i+v_i) \right) \:,
    \\
        V_{\alpha_i+\alpha_j, \beta_i+\beta_j}(\sigma_i,\sigma_j)&={\bf \Gamma}(-i(\alpha_i+\alpha_j) +x_i+ x_j,-i(\beta_i+\beta_j) +v_i+ v_j )
        \:.\label{dual4}
    \end{aligned}
\end{align}

The self-interaction term appearing in (\ref{stsqr}) is
\begin{align}
    S(\sigma_0)=x_0^2+v_0^2\:.\label{self4}
\end{align}


\section{Generalized star-triangle}

\begin{figure}[tbh]
\centering
\begin{tikzpicture}[scale=2]

\draw[-,thick] (-2,0)--(-2,1);
\draw[-,thick] (-2,0)--(-2.87,-0.5);
\draw[-,thick] (-2,0)--(-1.13,-0.5);

\filldraw[fill=black,draw=black] (-2,0) circle (1.2pt)
node[right=1.5pt] {\color{black} $\sigma_0$};

\filldraw[fill=black,draw=black] (-2,1) circle (1.2pt)
node[above=1.5pt] {\color{black} $\sigma_1$};
\filldraw[fill=black,draw=black] (-2.87,-0.5) circle (1.2pt)
node[left=1.5pt] {\color{black} $\sigma_3$};
\filldraw[fill=black,draw=black] (-1.13,-0.5) circle (1.2pt)
node[right=1.5pt] {\color{black} $\sigma_2$};

\fill[white!] (0.05,0.3) circle (0.01pt)
node[left=0.05pt] {\color{black}$=$};

\draw[-,dashed] (2,1)--(1.13,-0.5);
\draw[-,dashed] (1.13,-0.5)--(2.87,-0.5);
\draw[-,dashed] (2.87,-0.5)--(2,1);

\draw[dotted,thick] (2.0,0.03) circle (1cm);

\filldraw[fill=black,draw=black] (2,1) circle (1.2pt)
node[above=1.5pt]{\color{black} $\sigma_1$};
\filldraw[fill=black,draw=black] (1.13,-0.5) circle (1.2pt)
node[left=1.5pt]{\color{black} $\sigma_3$};
\filldraw[fill=black,draw=black] (2.87,-0.5) circle (1.2pt)
node[right=1.5pt]{\color{black} $\sigma_2$};

\end{tikzpicture}
\caption{The generalized star-triangle relation.}
\label{GSTR}
\end{figure}
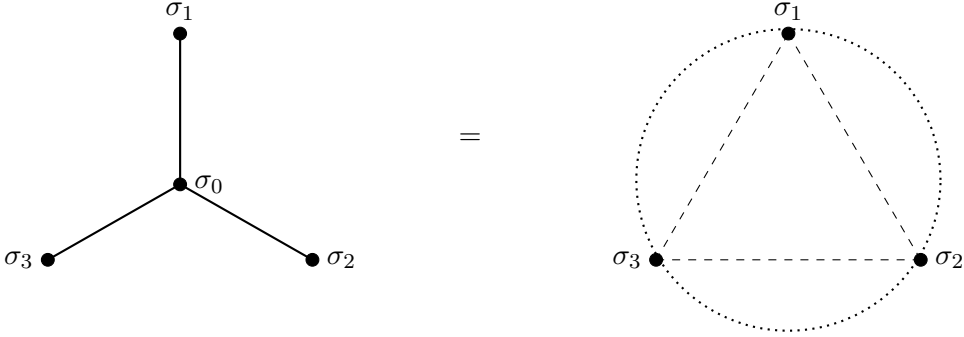

In this section, we study the generalized star-triangle relation \cite{Fisher1959, wegnerduality, Stre_ka_2010} which reveals the duality of lattice spin models. The duality appears between the nearest neighbor interacting model and a higher-spin interacting lattice spin model by showing the equality of their partition functions up to some coefficient.
The generalized star-triangle relation depicted in Fig.\ref{GSTR} consists of three nearest neighbor interactions at the left and four interactions which are three nearest neighbors (dashed lines) and one triple interaction (broken circle) at the right.

The generalized star-triangle relation with the Boltzmann weights is
\begin{align}
     \sum_{v_0}  \int dx_0 \: S(\sigma_0)\:&\prod_{i=1}^3 \:W_{\alpha_i,\beta_i}(\sigma_i,\sigma_0)   
\nonumber \\
&=\overline{V}_{\sum_{i=1}^3\alpha_i,\sum_{i=1}^3\beta_i}(\sigma_1,\sigma_2,\sigma_3) \prod_{1\leq i<j\leq 3} 
V_{\alpha_i+\alpha_j,\beta_i+\beta_j}(\sigma_i,\sigma_j)  \prod_{i=1}^3\overline{S}(\sigma_i)
   \label{str}\:,
\end{align}
where $\overline{V}_{\sum_{i=1}^3\alpha_i,\sum_{i=1}^3\beta_i}(\sigma_1,\sigma_2,\sigma_3)$ is the Boltzmann weight for the triple interaction and $\overline{S}$ is self-interaction term appearing at the right-hand-side due to non-uniform effective external field \cite{Stre_ka_2010}.

\subsection{Hyperbolic model}

In the integral identity (\ref{intssr}), we fix\footnote{One can see similar reductions in \cite{Spiridonov:2011hf}.} $ a_4 =\frac{\omega_1+\omega_2}{2}$ to obtain the generalized star-triangle relation \cite{wegnerduality}.

After the reduction, the integral identity (\ref{intssr}) becomes 
\begin{align}
\begin{aligned}
    \int_{-i\infty}^{i\infty}&\frac{\prod_{j=1}^3\gamma^{(2)}(a_j\pm z;\omega_1,\omega_2)}{\gamma^{(2)}(\pm2z;\omega_1,\omega_2)}\frac{dz}{2i\sqrt{\omega_1\omega_2}}
     \\    =&\gamma^{(2)}\left(\frac{\omega_1+\omega_2}{2}-\sum_{i=1}^3a_i ;\omega_1,\omega_2\right)\prod_{1\leq i<j\leq 3}\gamma^{(2)}(a_i+a_j;\omega_1,\omega_2)
    \\&\times \prod_{i=1}^3\gamma^{(2)}\left(a_i+\frac{\omega_1+\omega_2}{2};\omega_1,\omega_2\right)\:.
    \label{int}
    \end{aligned}
\end{align}
If we use the change of variables (\ref{chngvs}) but $j=\overline{1,3}$, we get Boltzmann weights of the model  defined in (\ref{integrableB1}) and external field (\ref{self1}) at the left-hand side and three-spin interaction, additional self-interactions together with two-spin interactions at the right-hand-side 
\begin{align}
    \begin{aligned}  
         \overline{V}_{\sum_{i=1}^3\alpha_i}(x_1, x_2, x_3 )&=\gamma^{(2)}\left(\frac{\omega_1+\omega_2}{2}+\sum_{i=1}^3(i\alpha_i -x_i) ;\omega_1,\omega_2\right)\:,
        \\
        V_{\alpha_i+\alpha_j}(x_i,x_j)&=\gamma^{(2)}(-i(\alpha_i+\alpha_j) +x_i+ x_j ;\omega_1,\omega_2)\:,
        \\
       \overline{S}(x_i) &=\gamma^{(2)}\left(-i\alpha_i +x_i+\frac{\omega_1+\omega_2}{2};\omega_1,\omega_2\right)\:.
    \end{aligned}
\end{align}

Like the star-square relation, the generalized star-triangle relation allows us to obtain another non-planar dual lattice spin model. In this hyperbolic solution to the generalized star-triangle relation, we acquire a novel model consisting of three-spin interactions and neighbor interactions as depicted in Fig.\ref{GSTR}.
\subsection{(Lens) Hyperbolic model}

The continuous fugacity will be fixed $ a_4 =\frac{\omega_1+\omega_2}{2}$ and also we need to fix the discrete flavor as $u_4=\frac{r}{2}$. Hence we obtain a new dual model to the Ising-like model that appeared in \cite{Gahramanov:2016ilb}. The reduced version of the integral identity (\ref{intssr2}) is
\begin{align} \nonumber
    \frac{1}{2r\sqrt{-\omega_1\omega_2}} &\sum_{y=0}^{[ r/2 ]}\epsilon (y) \int _{-\infty}^{\infty} \frac{\prod_{i=1}^3\gamma_h(a_i\pm z,u_i\pm y;\omega_1,\omega_2)}{\gamma_h(\pm 2z,\pm 2y;\omega_1,\omega_2)}  dz
     \nonumber \\
     =&
          \gamma_h\left(\frac{\omega_1+\omega_2}{2}-\sum_{i=1}^3a_i,-\frac{r}{2}-\sum_{i=1}^3u_i;\omega_1,\omega_2\right)
          \prod_{1\leq i<j\leq 3}\gamma_h(a_i + a_j,u_i + u_j;\omega_1,\omega_2)
          \nonumber\\& \times
          \prod_{i=1}^3   
          \gamma_h(\frac{\omega_1+\omega_2}{2}+a_i,\frac{r}{2}+u_i;\omega_1,\omega_2)\; .
\end{align}
The Boltzmann weight (\ref{integrableB2}) and the self-interaction term (\ref{self2}) at the right-hand side stay the same and the Boltzmann weight of three-spin interaction and nearest neighbor interactions of the dual model take the following form after the change of variable as (\ref{chngvs2}) but $j=1,2,3$ 
\begin{align}
    \begin{aligned}
\overline{V}_{\sum_{i=1}^3\alpha_i,\sum_{i=1}^3\beta_i}(\sigma_1,\sigma_2,\sigma_3)&=\gamma_h\left(\frac{\omega_1+\omega_2}{2}+\sum_{i=1}^3(i\alpha_i -x_i) ,-\frac{r}{2}+\sum_{i=1}^3(i\beta_i-v_i) ;\omega_1,\omega_2\right) \:,
        \\
        V_{\alpha_i+\alpha_j, \beta_i+\beta_j}(\sigma_i,\sigma_j)&=\gamma_h(-i(\alpha_i+\alpha_j) +x_i+ x_j,-i(\beta_i+\beta_j) +v_i+ v_j ;\omega_1,\omega_2)\:,
    \end{aligned}
\end{align}
where $\sigma=(x,v)$ represents both continuous and discrete spin variables, and the self-interaction term for the non-uniform external field is
\begin{align}
    \overline{S}(\sigma_i)=\gamma_h(\frac{\omega_1+\omega_2}{2}-i\alpha_i +x_i,\frac{r}{2}-i\beta_i +v_i;\omega_1,\omega_2)\:.
\end{align}


\subsection{Trigonometric model}
In this section, we reduce the identity (\ref{s2s1int}) by fixing the discrete parameter $u_4=0$ together with $a_4=q^{1/2}$ to
\begin{align}  \sum_{y\in\mathbb{Z}}\oint \frac{dz}{4 \pi i z}\frac{(1-q^{y} z^2)(1-q^{y} z^{-2})}{q^y     z^{4y} }
\frac{(q^{\frac{1}{2}+\frac{y}{2} }/{z},q^{\frac{1}{2}-\frac{y}{2} }{z};q)_\infty}{(q^{\frac{1}{2}+\frac{y}{2} }z,q^{\frac{1}{2}-\frac{y}{2} }/{z};q)_\infty}
\prod_{i=1}^3\frac{(q^{1+\frac{u_i+y}{2} }/{a_iz},q^{1+\frac{u_i-y}{2} }{z}/{a_i};q)_\infty}{(q^{\frac{u_i+y}{2} }a_iz,q^{\frac{u_i-y}{2} }{a_i}/{z};q)_\infty}\nonumber  \\\quad
= \frac{( q^{\frac{1}{2}+\frac{\sum_{i=1}^3 u_i}{2}}a_1 a_2 a_3 )_{\infty}}{(q^{\frac{1}{2}+ \frac{\sum_{i=1}^3 u_i}{2}}/a_1 a_2 a_3 )_{\infty}} \prod_{1\leq i<j\leq 3}  \frac{(q^{1+\frac{u_i+u_j}{2}}/a_ia_j;q)_\infty}{(q^{\frac{u_i+u_j}{2}}a_ia_j;q)_\infty}   \prod_{i=1}^3  \frac{(q^{\frac{1}{2}+\frac{u_i}{2}}/a_i;q)_\infty}{(q^{\frac{1}{2}+\frac{u_i}{2}}a_i;q)_\infty}.\end{align}

The change of variables $a_j=\alpha_j^{-1}x_{j} $ and $u_j=v_j-\beta_j$ where $j=1,2,3$ lead to the Boltzmann weights of the integrable model (\ref{integrableB3}) and the self-interaction term (\ref{self3}) for the right-hand side together with the following Boltzmann weight of three-spin interaction and neighbor interactions of dual lattice model
\begin{align}
 \begin{aligned}
    \overline{V}_{\sum_{i=1}^3\alpha_i,\sum_{i=1}^3\beta_i}(\sigma_1,\sigma_2,\sigma_3,\sigma_4)&=
        \frac{( q^{\frac{\sum_{i=1}^3(-\beta_i+v_i)}{2}}\prod_{i=1}^3\alpha_i^{-1}x_i;q)_{\infty}}{(q^{1+ \frac{\sum_{i=1}^3(-\beta_i+v_i)}{2}}\prod_{i=1}^3(\alpha_i^{-1} x_i)^{-1};q)_{\infty}} \:,
        \\
        V_{\alpha_i+\alpha_j, \beta_i+\beta_j}(\sigma_i,\sigma_j)&=\frac{(q^{1+\frac{\beta_i+\beta_j +v_i+ v_j}{2}}(\alpha_i^{-1}\alpha_j^{-1} x_i x_j)^{-1};q)_\infty}
{(q^{\frac{\beta_i+\beta_j +v_i+ v_j}{2}}\alpha_i^{-1}\alpha_j^{-1} x_i x_j;q)_\infty} 
 \:.
    \end{aligned}
\end{align}

After fixing parameters $a_4$ and $u_4$ in the integral identity (\ref{s2s1int}), there appears self-interaction $\overline{S}(\sigma_i)$ at the right-hand-side like the models mentioned before. However, in this model, there also exists an additional contribution\footnote{The main reason for this difference is the reflection property of the q-Pochammer symbol which is   
\begin{align}
    (z,q)_{\infty}(zq^{-1};q^{-1})_{\infty}=1\:.
\end{align} } to the self-interaction term at the left-hand side, i.e. $S(\sigma_0) \to \overline{\overline{S}}(\sigma_0)$ in the generalized star-triangle relation (\ref{str})
\begin{align}
    \overline{S}(\sigma_i)=\frac{(q^{\frac{1}{2}+\frac{(-\beta_i+v_i)}{2}}(\alpha_i^{-1} x_i)^{-1};q)_\infty}
{(q^{\frac{1}{2}+\frac{(-\beta_i+v_i)}{2}}\alpha_i^{-1} x_i;q)_\infty}\:,\quad \&\quad \overline{\overline{S}}(\sigma_0)=S(\sigma_0)\frac{\overline{S}(\sigma_0)}{\overline{S}(\widetilde{\sigma}_0)}\:,
\end{align}
where $\widetilde{\sigma}_0=(x_0,-v_0)$ and $\overline{\overline{S}}(\sigma_0)$ is a new self-interaction term at the left-hand-side but for the case of $m=0$, $\overline{\overline{S}}(\sigma_0)=S(\sigma_0)$ as in (\ref{self3}).

Once again, we obtain a new dual lattice spin model by solving the generalized star-triangle relation.

\subsection{Rational model}

We set variables as $a_4=-i$ and $u_4=0$ and use the properties of the gamma function  (\ref{gamma}) to obtain the following identity

\begin{align} \begin{aligned}
    \sum_{y\in {\mathbb Z}}&\int_{-\infty}^{\infty}\big(z^2+y^2\big)\prod_{i=1}^3{\bf \Gamma}(a_i\pm z,u_i\pm y)\frac{dz}{ 8\pi}
\\ \qquad{}
&= {\bf \Gamma}\Big(-i-\sum\limits_{i=1}^ 3 a_i,\sum\limits_{i=1}^ 3 u_k\Big)
\prod\limits_{1\leq i < j \leq 3}
{\bf \Gamma}(a_i+a_j,u_i+u_j)\prod\limits_{i=1}^3
{\bf \Gamma}(a_i-i,u_i)\:.
\end{aligned}
\end{align}

The change of variables in (\ref{chngvs2}) is applied to obtain the Boltzmann weights of the integrable model (\ref{integrableB4}) and the three-spin interaction of the dual non-planar lattice model
\begin{align}
    \begin{aligned}
\widetilde{V}_{\sum_{i=1}^3\alpha_i,\sum_{i=1}^3\beta_i}(\sigma_1,\sigma_2,\sigma_3)&={\bf \Gamma}\left(-i-\sum_{i=1}^3(-i\alpha_i +x_i) ,\sum_{i=1}^3(-i\beta_i+v_i) \right) \:,
\\        V_{\alpha_i+\alpha_j, \beta_i+\beta_j}(\sigma_i,\sigma_j)&={\bf \Gamma}(-i(\alpha_i+\alpha_j) +x_i+ x_j,-i(\beta_i+\beta_j) +v_i+ v_j )\:,
    \end{aligned}
\end{align}
 and the self-interaction term on the right-hand side is
\begin{align}
    \overline{S}(\sigma_i)={\bf \Gamma}(-i\alpha_i +x_i-i,v_i)\:.
\end{align}

Then one can observe that Boltzmann weights satisfy the generalized star-triangle relation. 

\section{Conclusion}

In this study, we acquire new lattice spin models consisting of higher-spin interactions for Ising-like models via the gauge/YBE correspondence. The novel non-planar lattice spin models are dual to the Ising-like models containing the nearest neighbor interactions.
These non-planar lattice models in statistical mechanics are obtained by solving the star-square relation and the generalized star-triangle relation. The duality transformations are studied by the integral identities coming from the equality of partition functions of the dual three-dimensional supersymmetric gauge theories

Investigating various solutions to the star-square relation and the generalized star-triangle relation via different dual three-dimensional supersymmetric gauge theories results in terms of hyperbolic, lens hyperbolic, trigonometric, and rational functions. Solving the star-square relation and the generalized star-triangle relation allows us to construct models containing four-spin and three-spin interactions, respectively. 

Besides various interesting connections between integrable models and supersymmetric gauge theories via the gauge/YBE correspondence, we observe that the correspondence has much more open directions to study links between supersymmetric gauge theories and lattice spin models in statistical mechanics. 

One of the further studies could be the investigation of the star-square relation for the integrable models such as lens elliptic model \cite{Yamazaki:2013nra}, lens hyperbolic model \cite{Gahramanov:2016ilb}, trigonometric model \cite{Gahramanov:2015cva} and rational model \cite{Kels:2013ola} obtained via the supersymmetric dualities.

It is also interesting to obtain star-triangle relation for the lattice spin models where spins interact solely with Boltzmann weight $V_{\alpha_i+\alpha_j, \beta_i+\beta_j}$ instead of mixing with $W_{\alpha_i+\alpha_j, \beta_i+\beta_j}$ as occurred in \cite{Kels:2018xge}. 


\section*{Acknowledgements}

It is a pleasure to thank Ilmar Gahramanov and Andrew P. Kels for the helpful discussions. We would like to thank Hesam Soltanpanahi for the warm hospitality at the Jagiellonian University (Krakow, Poland), where most of this work was completed.  Mustafa Mullahasanoglu is supported by the 3501-TUBITAK
Career Development Program under grant number 122F451.



\bibliographystyle{utphys}
\bibliography{refYBE}

\end{document}